\documentclass[aps,twocolumn,groupedaddress]{revtex4}
\usepackage{graphicx}

\begin{document}

\title{Chern Numbers for Spin Models of Transition Metal Nanomagnets}


\author{C.M.~Canali$^{1}$ A.~Cehovin$^{2}$ and A.H.~MacDonald$^{3}$}
\affiliation{$^1$Department of Technology, Kalmar University, 391 82 Kalmar,
Sweden}
\affiliation{$^2$Division of Solid State Theory, Department of Physics,
Lund University, SE-223 62 Lund, Sweden}
\affiliation{$^3$Department of Physics, University of Texas at Austin,
Austin TX 78712}


\date{\today}

\begin{abstract}

We argue that ferromagnetic transition metal nanoparticles  
with fewer than approximately 100 atoms can be described by 
an effective Hamiltonian with a single giant spin degree of freedom.
The total spin $S$ of the effective Hamiltonian is specified by a  
Berry curvature Chern number that characterizes the topologically
non-trivial dependence of a nanoparticle's many-electron wavefunction on
magnetization orientation.  The Berry curvatures and associated Chern numbers
have a complex dependence on spin-orbit coupling in the nanoparticle 
and influence the semiclassical Landau-Liftshitz equations that describe 
magnetization orientation dynamics.

\end{abstract}

\maketitle


Both molecular nanomagnets\cite{sessoli1993,friedman1996,wernsdorfer2001}
and ferromagnetic transition metal
clusters\cite{billas1994, lederman1994, wernsdorfer97a} have been actively 
studied over the past 
decade.  Interest in molecular nanomagnets has been due primarily to their 
position near the borderline between quantum and classical 
behaviors\cite{thomas1996, qtm94,wernsdorfer2001}.  
For ferromagnetic transition metal nanoparticles, on the other hand, interest 
has been spurred mainly by classical physics issues relevant to information 
storage\cite{majetich1999, murray_science2000,jamet2001} and by the 
interplay between collective and quasiparticle degrees
of freedom\cite{gueron1999,deshmukh2001}.  The present work 
is motivated by the observation that the low energy physics of small transition metal
clusters can be described by an effective Hamiltonian
with a single giant spin degree of freedom, like that of a molecular magnet.
A transition metal nanoparticle will behave like a 
molecular magnet when the energy scale associated with its collective
magnetization orientation, the anisotropy energy,
does not exceed the smallest energy scale associated with its quasiparticle degrees-of-freedom,
the single-particle level spacing $\delta$.  
When bulk density-of-states and anisotropy
energy values are used to estimate the particle-size at which this condition
is satisfied, the cubic transition metal ferromagnets 
Fe and Ni are predicted to act like 
molecular magnets when the number of atoms $N_A$ is smaller than $\sim 1000$,
while Co is predicted to act like a molecular magnet for $N_A$ smaller than $\sim 100$. (See Table \ref{table1}).
It is remarkable that an extensive ($\propto N_A$) 
energy scale, like the total anisotropy energy, is smaller than a microscopic energy 
scale ($\propto N_A^{-1}$), like the level spacing, at a relatively large 
particle number.  This surprising property arises from a combination of relatively weak
spin-orbit coupling in the 3d transition-metal series and the itinerant character
of transition metal ferromagnetism.  The magnetocrystalline anisotropy energy in these systems 
in the bulk is five to six orders of magnitude smaller than the magnetic
condensation energy. 
Even accounting for the substantially larger anisotropy energy per volume expected in 
typically shaped nanoparticles, it seems clear that the total width of the anisotropy energy
landscape will often be smaller than the level spacing for $N_A$ smaller than $\sim  100$.  
The physics that controls the Hamiltonian and, most centrally,
the total spin of these malleable molecular magnets is the subject of this Letter.

In the absence of spin-orbit and dipole-dipole interactions, the collective physics of a 
transition metal nanoparticle is easily understood\cite{cmc_ahm2000prl}.
Quasiparticle states, calculated
for a particular magnetization orientation (using the spin-density-functional
theory recipe for example) have definite spin-orientations.  The many-particle
ground state has a total spin quantum number 
$S=(N_{a}-N_{o})/2$, where $N_{a}$ and $N_{o}$ are 
respectively the number of quasiparticle states whose individual 
spins are {\it aligned
with} and {\it opposed to} the total spin.  
The $2S+1$ fold ground-state degeneracy is lifted by 
an arbitrarily weak external magnetic field, selecting a ground state with 
magnetic moment $2 S \mu_B$.  This simplicity dissolves into 
a complex muddle when spin-orbit interactions are included.  
Individual quasiparticle
states no-longer have definite spin orientations.  It is not immediately
obvious how to derive an effective Hamiltonian which describes the low-energy
collective physics, or even how to determine the dimension of the quantum
Hilbert space for the coherent magnetization degree of freedom.  
We address this issue here using a path integral approach.  
We find that the total 
spin $S$ of the molecular magnet is 
specified by a Chern number defined by the dependence of
the many-particle wavefunction phase on magnetization orientation,
and propose a procedure for extracting a quantum Hamiltonian from the
dependence of energy and Berry curvature on orientation.
\begin{table}
\caption{
$\delta\,N_A$ is the bulk mean-level spacing times the number of atoms.
$E_B$ is the bulk coherent rotation anisotropy energy barrier
(For cubic systems $E_B/{\rm vol}$ is one third of magnetic anisotropy 
constant $K_1$, whereas in the uniaxial case it is equal to the magnetic anisotropy constant.) 
$N_A^{\star}$ is the number of atoms at which $\delta$ and $E_B$
become equal.
\label{table1}}
\begin{ruledtabular}
\begin{tabular}{cccccc}
 &$\delta\,\,N_A$ (meV)\footnotemark[1]&
$E_B/{\rm vol}$ (MJ/m$^3$)\footnotemark[2]&
 $E_B/N_A$ (meV) &$N_A^{\star}$\\
\colrule
Fe& 695 &  0.016 &.00118
& 767 \\
Co& 581 &  0.53 &.0367
& 126 \\
Ni& 495 &  0.002 &.000114
& 2080 \\
\end{tabular}
\end{ruledtabular}
\footnotetext[1]{From Ref.~\onlinecite{papaconstantopoulos}.}
\footnotetext[2]{Bulk value at room temperature from Ref.~\onlinecite{skomski}.}
\end{table}
Our analysis is based on an approximate imaginary-time quantum action with a 
single magnetization-orientation degree of freedom, 
$\hat n(\tau)$:
\begin{equation}
{\cal S}_{\rm coh}[\hat n] \equiv \int d \tau  
\Big [\Big \langle \Psi[\hat n] \Big \vert \vec \nabla_{\hat n}\Psi[\hat n]\Big \rangle\cdot
{\partial \hat n \over \partial \tau} +  E[\hat n] \Big]\;.
\label{action}
\end{equation}
For simple model Hamiltonians in which the interaction terms can be represented exactly by 
auxiliary field function integrals\cite{fradkin}, 
actions of this form can be derived fully
microscopically\cite{cmc_ac_ahm2002pap4}.  For the realistic 
description of real nanoparticles, however, this action is approximate 
and should be constructed by using the family of spin-density-functional theory 
constrained Kohn-Sham states with net magnetization orientation $\hat n$
for an approximate identity resolution at energy scales below the single-particle level spacing.
In the absence of spin-orbit interactions, spin-density-functional theory 
describes the ground state in terms of Kohn-Sham quasiparticles that experience 
a scalar potential and an exchange-correlation 
effective magnetic field\cite{andropov1996}, 
both of which have a complex dependence on 
spatial coordinate.  The orientational degree of freedom we have in mind
for the action is the direction $\hat n$
in spin-space of the spin-density-functional theory 
exchange-correlation effective field\cite{uhl1994}, or its spatial average if 
low-energy states have non-collinear magnetization. 
In Eq.[~\ref{action}] $\vert \Psi [\hat n] \rangle$ is the 
Kohn-Sham single-Slater determinant state defined by this orientation, with  
spin-orbit terms included in the Hamiltonian and 
$E[\hat n]$ is the density-functional-theory energy.  
The first term in this action is a Berry phase contribution\cite{resta2000} 
whose role in 
quantizing small-amplitude magnetic orientation fluctuations (spin-waves) 
of bulk transition metal ferromagnets 
has been successfully exploited by 
Niu {\it et.al.}\cite{niu1998, niu1999, bylander2000},
and other authors\cite{gebauer2000}.
When spin-orbit interactions 
are included, both Berry phase and
energy function $E[\hat n]$ terms will have a non-trivial 
dependence on orientation{\cite{ahm_cmc2001ssc, cmc_ac_ahm2002pap4}.
We expect that quantitatively accurate actions can be constructed in this way
for completely specified nanoparticles; in the rest of this paper we discuss 
some qualitative properties using a 
tight-binding model\cite{ac_cmc_ahm2002}.

It is useful to express the Berry phase 
in a gauge invariant form\cite{auerbach,resta2000}.
During its imaginary-time evolution, the unit vector $\hat n$ traces 
a closed trajectory on the unit sphere. By Stokes' theorem we can
rewrite the closed line integral and express the Berry phase in terms of  
a surface integral over the area enclosed by the path
\begin{equation}
\label{berry_loop}
{\cal S}_{\rm Berry}[\hat n] =\,\oint d \hat n \cdot 
\Big\langle \Psi \vert {\vec \nabla_{\hat n} \Psi}
\Big\rangle\ = \,\int_{\rm area} 
\vec \nabla_{\hat n} \times\Big\langle \Psi \vert
 \vec \nabla_{\hat n} \Psi \Big\rangle \cdot \hat n\, d a\;.
\end{equation}
The Berry curvature  
\begin{equation}
\label{curvature}
{\cal C}[\hat n] \equiv i\,\vec \nabla_{\hat n} \times 
\Big \langle \Psi[\hat n]\Big\vert
 \vec \nabla_{\hat n}\Psi[\hat n]\Big  \rangle \cdot \hat n\;,
\end{equation}
is gauge invariant and the Chern number $S$, 
defined as the average of the curvature 
over the unit sphere, is required to be half of an integer\cite{simon1983}.   
The Chern number appears below as the total spin in the effective model for
the nanoparticle, hence the notation chosen above. 

To discuss the Chern numbers, it is useful to start 
from the case of no spin-orbit coupling, 
for which $S =(N_{a}-N_{o})/2$.
To see this, consider the Kohn-Sham single-Slater determinant many-particle state
which describes magnetization in direction $\hat n$: 
\begin{equation}
\Big |\Psi[\hat n]\Big \rangle = \prod_i\Big[u(\hat n)c^{\dagger}_{i\uparrow} +
v(\hat n)c^{\dagger}_{i\downarrow}\Big]\,
\prod_j \Big[-v^{\star}(\hat n)c^{\dagger}_{j\uparrow} +
u(\hat n)c^{\dagger}_{j\downarrow}\Big]|0\rangle\;, 
\end{equation}
where $c^{\dagger}_{\sigma}$ 's are electron creation operators 
defined with the arbitrary polar angle of the coordinate system
used to measure magnetization orientations as the spin-quantization axis,
and $i$ and $j$ are orbital indices running respectively 
over $N_{a}$ majority-spin orbitals
and $N_{o}$ minority-spin orbitals.  We have made the 
gauge choice 
$u(\hat n)=\cos(\theta/2)$, $v(\hat n)= e^{-i\varphi}\sin(\theta/2)$
for the $S=1/2$ quasiparticle coherent states,  
where $\theta$ and $\varphi$ are the polar and azimuthal angles 
used to specify $\hat n(\theta,\varphi)$.
Given this state, 
the Berry connection\cite{auerbach,resta2000},
$\vec A[\hat n] \equiv
i\Big\langle\Psi[\hat n]\Big|\vec\nabla_{\hat n}\Psi[\hat n]\Big\rangle$,
is easily calculated
\begin{eqnarray}
\vec A[\hat n] = 
i\sum_i\Big[ u {\partial u \over  \partial \hat n}
+ v^{\star}{\partial v \over  \partial \hat n}\Big] 
+ i\sum_j\Big[ v {\partial v^{\star} \over  \partial \hat n} 
+ u {\partial u \over  \partial \hat n}\Big]\nonumber\\
=\Bigg(\sum_i 1-\sum_j 1\Bigg) \;{1-\cos(\theta)\over 2\sin(\theta)}\hat \varphi
= \,{N_a - N_o\over 2}\,
{1-\cos(\theta)\over \sin(\theta)}\,\hat \varphi\;,
\end{eqnarray}
It follows that the Berry curvature 
${\cal C}[\hat n]= \vec \nabla_{\hat n} \times\vec A[\hat n]$
is constant and equal to 
\begin{equation}
{\cal C} =
{N_a - N_o\over 2}\;.
\end{equation}
In this case each majority-spin Kohn-Sham quasiparticle contributes 
$1/2$ to the Chern number while each minority-spin quasiparticle 
contributes $-1/2$.

When the spin-orbit interaction is present, the Berry curvature is no
longer a constant, but the Chern number $S$ must still be half-integer. 
The total Chern number $S$ of the nanoparticle
can still be considered as the sum over contributions from each 
quasiparticle state\cite{niu1998}.  Individual quasiparticle Chern number
properties are familiar from related studies, especially in connection
with the topological interpretation of the quantized 
Hall conductance in a two-dimensional 
electron gas\cite{tkn2,avron1983,simon1983,kohmoto1985}.
They can change only when quasiparticle level degeneracies 
occur. In nanoparticles without special symmetries, level degeneracies
will occur at discrete values of $\hat n$, 
and at isolated values of any microscopic Hamiltonian parameter.
When two levels intersect and then come apart again as a parameter is
varied, then their individual Chern numbers may not be conserved but their
sum is conserved\cite{avron1983}.
As we now show,
spin-orbit interactions lead to surprisingly ubiquitous 
and large deviations of quasiparticle 
Chern number values from their $\pm 1/2$ values in the 
absence of spin-orbit coupling.
%
%
%
 \begin{figure}
 \includegraphics[width=2.4in,height=2.9in]{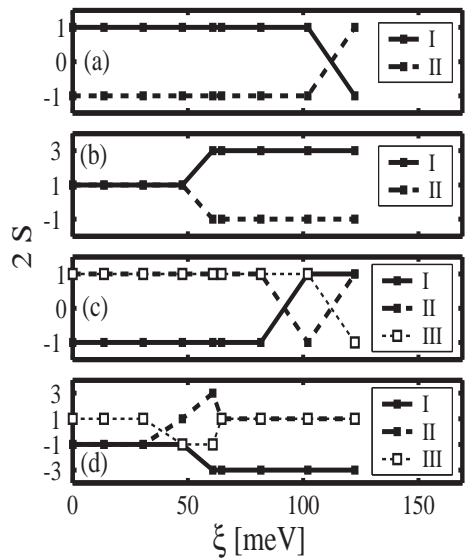}
 \caption{Chern numbers of quasiparticle energy levels
of a 25-atom Cobalt nanoparticle,
as a function of the spin-orbit coupling strength $\xi$. The nanoparticle
is modeled by a tight-binding Hamiltonian\cite{ac_cmc_ahm2002}.
In each panel I, II, (and sometimes III) label two (or three) contiguous
orbitals whose energies are ordered in ascending order.
Deviations of individual $S$ from their zero spin-orbit coupling values $\pm 1/2$
occur when two levels cross.
Note that the sum of the two Chern numbers
is conserved at crossing.}
 \label{fig1}
 \end{figure}

 \begin{figure}
\includegraphics[width=2.6in,height=2.6in]{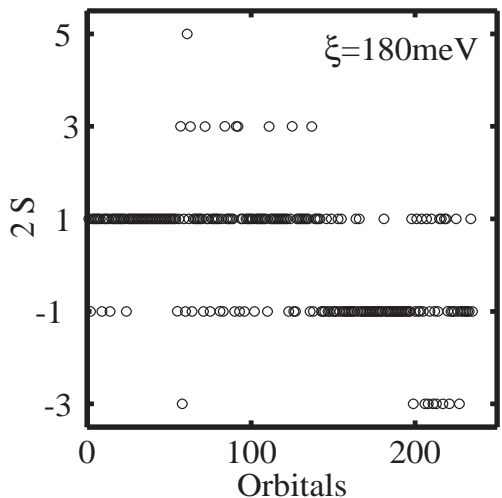}
 \caption{Chern numbers of all the individual quasiparticle energy levels
of a 25-atom nanoparticle
for a given value of $\xi$.
As a result of repeated level crossings, several of these numbers differ
strongly from $\pm 1/2$.}

 \label{fig2}
 \end{figure}

 \begin{figure}
\includegraphics[width=3.2in,height=2.3in]{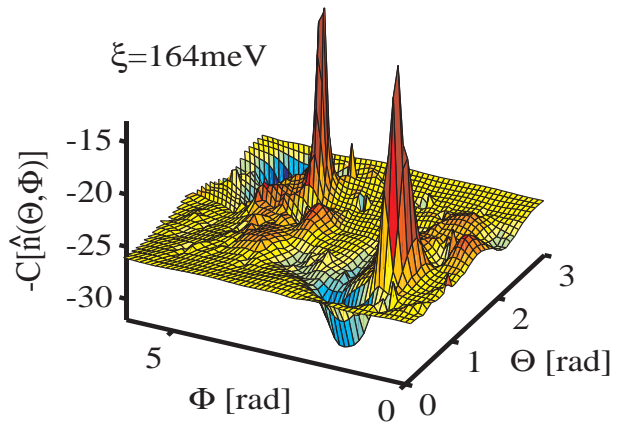}
 \caption{Planar projection of the total (negative) Berry curvature $-{\cal C}[\hat n]$ for
a 25-atom nanoparticle. The strength of the spin-orbit
coupling $\xi= 165$ meV
is chosen in proximity to a value at which the Fermi level and the
first unoccupied level cross. Level crossing occurs
for a given $\hat n$ and its opposite $-\hat n$. 
In correspondence of these two values,
$-{\cal C}[\hat n]$ has large peaks.}
 \label{fig3}
 \end{figure}
%
%
In Fig.~\ref{fig1} we plot the variation of quasiparticle Chern numbers
for representative groups of contiguous levels 
of a nanoparticle modeled by a tight-binding model \cite{ac_cmc_ahm2002},
as a function of the spin-orbit coupling
strength $\xi$. Typically we find that
individual Chern numbers of pairs of levels undergoing level crossing
experience a change of $\pm 1$, with their sum remaining unchanged.
These changes can be thought of as representing changes in the {\em orbital} contribution
to the effective angular momentum of individual orbitals. 
Note that level crossings always occur in pairs, for a given value
of $\hat n$ and its opposite $-\hat n$\footnote{This is due to the
degeneracy of the system with respect to $\hat n \to -\hat n$, which
is rooted in the time reversal symmetry of the system in absence of an
external field.}.  
As a result of repeated level crossing, some of the
individual Chern numbers end up with values very different from $\pm 1/2$
as shown in Fig.~\ref{fig2}, where we plot all individual
quasiparticle Chern numbers for a given $\xi$.
%
%
If a level crossing occurs between the Fermi level and the first
unoccupied level, the {\it total} Chern number will change from
the original zero spin-orbit value $(N_a - N_{o})/ 2$. 
In the vicinity of such a level crossing, the Berry curvature ${\cal C}[\hat n]$
will deviate strongly from its average value as a function of $[\hat n]$.
An example of this is shown in Fig.~\ref{fig3},
where we plot ${\cal C}[\hat n]$ at a value of $\xi$ for which
the Fermi level and the first unoccupied level are almost degenerate
in two directions of $\hat n$ and the Berry curvature has sharp peaks.

Assuming only that the Berry curvature is positive-definite, 
our low-energy effective
action can be mapped to that of a quantum spin Hamiltonian by
making a change of variables that transforms the Berry curvature 
to a constant.  The  
Hamiltonian representation is easier to use for 
explicit calculations of collective tunneling amplitudes, 
non-linear response to 
external electromagnetic fields and other relevant properties.
To be explicit, we change variables from $u=cos(\theta)$ and 
$\varphi$ to $u'$ and $\varphi'$ defined by
\begin{eqnarray}
\phi' &=&  \frac{2 \pi \int_{0}^{\phi} d \varphi'' {\cal C}(u,\varphi'')}
{\int_{0}^{2\pi } d \varphi'' {\cal C}(u,\varphi'')}\\
u' &=& -1 +  \frac{1}{2 \pi S}\int_{-1}^{u} du'' \int_{0}^{2\pi} d \varphi''
{\cal C}(u'',\varphi'') 
\end{eqnarray} 
With this change of variables ${\cal C}(u,\varphi)\;du d\varphi  = 
S\; du' d \varphi'$ and the real-time action for a path can be written 
\begin{equation}
{\cal S}_{\rm spin}[\hat n'] =
\int_0^t dt'\; \Big (\vec A\cdot d\hat n'/dt' - 
{S\over {\cal C}\big[\hat n\big(\hat n'(t')\big)\big]}\,E\big[\hat n\big(\hat n'(t')\big)\big]
\Big)\,,
\label{action_spin}
\end{equation}
where $\vec A = S\big (1-\cos(\theta')\big)/\sin(\theta')\hat \phi'$. 
This is the 
quantum action for a total spin quantum number $S$ with the {\em classical
Hamiltonian}\cite{auerbach} specified by the 
second term of the integrand, {\it i.e.} 
\begin{equation}
H[\hat n'(t')] \equiv \langle S, \hat n'(t')| {\cal H}|S,\hat  n'(t')\rangle
\equiv 
{S\, E\big[\hat n\big(\hat n'(t')\big)\big]
\over {\cal C}\big[\hat n\big(\hat n'(t')\big)\big]}\:
\end{equation}
where ${\cal H}$ is the quantum Hamiltonian of the spin system and
$|S, \hat n'(t')\rangle $ is a spin-$S$ coherent state parametrized by
the unit vector $\hat n'(t')$.  Given the energy and 
Berry curvature functions, this quantum Hamiltonian can always be explicitly
constructed. Applications of this procedure will be presented elsewhere.

A non-constant Berry curvature affects the 
Landau-Lifshitz equations describing the precession motion of
the nanoparticle magnetic moments.
These equations are equivalent to the Euler-Lagrange equations of motion
derived from the real-time action ${\cal S}_{\rm spin}[\hat n']$
in the semiclassical (large $S$) approximation:
\begin{equation}
\dot {\hat n´}_{\rm cl}(t')= \hat n´_{\rm cl}(t')\times 
{\partial \Big(H[\hat n'(t')]/S\Big)\over \partial \hat n'(t')}\Bigg\vert_{\hat n´_{\rm cl}}\:,
\label{LL}
\end{equation}
where ${\hat n´}_{\rm cl}(t')$ is determined by the saddle-point equations
${\delta {\cal S}_{\rm spin}[\hat n'(t')]\over \delta \hat n'(t')}
\Big\vert_{\hat n´_{\rm cl}}=0$.
Using the expression of $H$ in Eq.~\ref{LL}, we can see that a non-constant
Berry curvature modifies the precession rate of the magnetic moment
fluctuations. This effect is particularly
important when the Fermi level and the first unoccupied state are close
to a degeneracy point for some value of $\hat n$, since as shown in
Fig~\ref{fig3}, ${\cal C}[\hat n]$ can then deviate considerably 
from the Chern number $S$.

In conclusion, we have derived an effective spin Hamiltonian with a
single giant spin degree of freedom that describes 
the low-energy physics of a small metallic nanomagnet. The dimension of the Hilbert
space of the effective Hamiltonian is given by a Berry curvature Chern number
which has a non-trivial dependence on spin-orbit coupling strength and nanoparticle
details.
We would like to thank G. Canright and W. Wersdorfer for helpful conversations.
This work was supported in part by the Swedish Research Council
under Grant No:621-2001-2357, by the faculty of natural sciences 
of Kalmar University, and by the National Science Foundation
under Grants DMR 0115947 and DMR 0210383.
\bibliography{./Biblio}

\begin{thebibliography}{32}
\expandafter\ifx\csname natexlab\endcsname\relax\def\natexlab#1{#1}\fi
\expandafter\ifx\csname bibnamefont\endcsname\relax
  \def\bibnamefont#1{#1}\fi
\expandafter\ifx\csname bibfnamefont\endcsname\relax
  \def\bibfnamefont#1{#1}\fi
\expandafter\ifx\csname citenamefont\endcsname\relax
  \def\citenamefont#1{#1}\fi
\expandafter\ifx\csname url\endcsname\relax
  \def\url#1{\texttt{#1}}\fi
\expandafter\ifx\csname urlprefix\endcsname\relax\def\urlprefix{URL }\fi
\providecommand{\bibinfo}[2]{#2}
\providecommand{\eprint}[2][]{\url{#2}}

\bibitem[{\citenamefont{Sessoli et~al.}(1993)\citenamefont{Sessoli, Gatteschi,
  Caneschi, and Novak}}]{sessoli1993}
\bibinfo{author}{\bibfnamefont{S.}~\bibnamefont{Sessoli}},
  \bibinfo{author}{\bibfnamefont{D.}~\bibnamefont{Gatteschi}},
  \bibinfo{author}{\bibfnamefont{A.}~\bibnamefont{Caneschi}}, \bibnamefont{and}
  \bibinfo{author}{\bibfnamefont{M.~A.} \bibnamefont{Novak}},
  \bibinfo{journal}{Nature} \textbf{\bibinfo{volume}{383}},
  \bibinfo{pages}{141} (\bibinfo{year}{1993}).

\bibitem[{\citenamefont{Friedman et~al.}(1996)\citenamefont{Friedman, Sarachik,
  and Tejada}}]{friedman1996}
\bibinfo{author}{\bibfnamefont{J.~R.} \bibnamefont{Friedman}},
  \bibinfo{author}{\bibfnamefont{M.~P.} \bibnamefont{Sarachik}},
  \bibnamefont{and} \bibinfo{author}{\bibfnamefont{J.}~\bibnamefont{Tejada}},
  \bibinfo{journal}{Phys. Rev. Lett.}
  \textbf{\bibinfo{volume}{76}}(\bibinfo{number}{20}), \bibinfo{pages}{3830}
  (\bibinfo{year}{1996}).

\bibitem[{\citenamefont{Wernsdorfer}(2002)}]{wernsdorfer2001}
\bibinfo{author}{\bibfnamefont{W.}~\bibnamefont{Wernsdorfer}},
  \bibinfo{journal}{Adv. Chem. Phys.} \textbf{\bibinfo{volume}{118}},
  \bibinfo{pages}{99} (\bibinfo{year}{2002}).

\bibitem[{\citenamefont{Billas et~al.}(1994)\citenamefont{Billas, Ch\^atelain,
  and de~Heer}}]{billas1994}
\bibinfo{author}{\bibfnamefont{I.~M.~L.} \bibnamefont{Billas}},
  \bibinfo{author}{\bibfnamefont{A.}~\bibnamefont{Ch\^atelain}},
  \bibnamefont{and} \bibinfo{author}{\bibfnamefont{W.~A.}
  \bibnamefont{de~Heer}}, \bibinfo{journal}{Science}
  \textbf{\bibinfo{volume}{265}}, \bibinfo{pages}{1682} (\bibinfo{year}{1994}).

\bibitem[{\citenamefont{Lederman et~al.}(1994)\citenamefont{Lederman, Shultz,
  and Ozachi}}]{lederman1994}
\bibinfo{author}{\bibfnamefont{M.}~\bibnamefont{Lederman}},
  \bibinfo{author}{\bibfnamefont{S.}~\bibnamefont{Shultz}}, \bibnamefont{and}
  \bibinfo{author}{\bibfnamefont{M.}~\bibnamefont{Ozachi}},
  \bibinfo{journal}{Phys. Rev. Lett.}
  \textbf{\bibinfo{volume}{73}}(\bibinfo{number}{14}), \bibinfo{pages}{1986}
  (\bibinfo{year}{1994}).

\bibitem[{\citenamefont{Wernsdorfer et~al.}(1997)\citenamefont{Wernsdorfer,
  Orozco, Hasselbach, Benoit, Demoncy, Loiseau, and Maillya}}]{wernsdorfer97a}
\bibinfo{author}{\bibfnamefont{W.}~\bibnamefont{Wernsdorfer}},
  \bibinfo{author}{\bibfnamefont{E.~B.} \bibnamefont{Orozco}},
  \bibinfo{author}{\bibfnamefont{K.}~\bibnamefont{Hasselbach}},
  \bibinfo{author}{\bibfnamefont{A.}~\bibnamefont{Benoit}},
  \bibinfo{author}{\bibfnamefont{N.}~\bibnamefont{Demoncy}},
  \bibinfo{author}{\bibfnamefont{A.}~\bibnamefont{Loiseau}}, \bibnamefont{and}
  \bibinfo{author}{\bibfnamefont{H.~P.~D.} \bibnamefont{Maillya}},
  \bibinfo{journal}{Phys. Rev. Lett.}
  \textbf{\bibinfo{volume}{78}}(\bibinfo{number}{9}), \bibinfo{pages}{1791}
  (\bibinfo{year}{1997}).

\bibitem[{\citenamefont{Thomas et~al.}(1996)\citenamefont{Thomas, Lionti,
  Ballou, Gatteschi, Sessoli, and Barbara}}]{thomas1996}
\bibinfo{author}{\bibfnamefont{L.}~\bibnamefont{Thomas}},
  \bibinfo{author}{\bibfnamefont{F.}~\bibnamefont{Lionti}},
  \bibinfo{author}{\bibfnamefont{R.}~\bibnamefont{Ballou}},
  \bibinfo{author}{\bibfnamefont{D.}~\bibnamefont{Gatteschi}},
  \bibinfo{author}{\bibfnamefont{S.}~\bibnamefont{Sessoli}}, \bibnamefont{and}
  \bibinfo{author}{\bibfnamefont{B.}~\bibnamefont{Barbara}},
  \bibinfo{journal}{Nature} \textbf{\bibinfo{volume}{383}},
  \bibinfo{pages}{145} (\bibinfo{year}{1996}).

\bibitem[{\citenamefont{Gunther and Barbara}(1995)}]{qtm94}
\bibinfo{editor}{\bibfnamefont{L.}~\bibnamefont{Gunther}} \bibnamefont{and}
  \bibinfo{editor}{\bibfnamefont{B.}~\bibnamefont{Barbara}}, eds.,
  \emph{\bibinfo{title}{Quantum Tunneling of Magnetization}}, QTM´94
  (\bibinfo{publisher}{Kluwer, Dordrecht}, \bibinfo{year}{1995}).

\bibitem[{\citenamefont{Majetich and Jin}(1999)}]{majetich1999}
\bibinfo{author}{\bibfnamefont{S.~A.} \bibnamefont{Majetich}} \bibnamefont{and}
  \bibinfo{author}{\bibfnamefont{Y.}~\bibnamefont{Jin}},
  \bibinfo{journal}{Science}
  \textbf{\bibinfo{volume}{284}}(\bibinfo{number}{5413}), \bibinfo{pages}{470}
  (\bibinfo{year}{1999}).

\bibitem[{\citenamefont{Sun et~al.}(2000)\citenamefont{Sun, Murray, Weller,
  Folks, and Moser}}]{murray_science2000}
\bibinfo{author}{\bibfnamefont{S.}~\bibnamefont{Sun}},
  \bibinfo{author}{\bibfnamefont{C.~B.} \bibnamefont{Murray}},
  \bibinfo{author}{\bibfnamefont{D.}~\bibnamefont{Weller}},
  \bibinfo{author}{\bibfnamefont{L.}~\bibnamefont{Folks}}, \bibnamefont{and}
  \bibinfo{author}{\bibfnamefont{A.}~\bibnamefont{Moser}},
  \bibinfo{journal}{Science}
  \textbf{\bibinfo{volume}{287}}(\bibinfo{number}{5460}), \bibinfo{pages}{1989}
  (\bibinfo{year}{2000}).

\bibitem[{\citenamefont{Jamet et~al.}(2001)\citenamefont{Jamet, Wernsdorfer,
  Thirion, Mailly, Dupuis, M\'elinon, and P\'eres}}]{jamet2001}
\bibinfo{author}{\bibfnamefont{M.}~\bibnamefont{Jamet}},
  \bibinfo{author}{\bibfnamefont{W.}~\bibnamefont{Wernsdorfer}},
  \bibinfo{author}{\bibfnamefont{C.}~\bibnamefont{Thirion}},
  \bibinfo{author}{\bibfnamefont{D.}~\bibnamefont{Mailly}},
  \bibinfo{author}{\bibfnamefont{V.}~\bibnamefont{Dupuis}},
  \bibinfo{author}{\bibfnamefont{P.}~\bibnamefont{M\'elinon}},
  \bibnamefont{and} \bibinfo{author}{\bibfnamefont{A.}~\bibnamefont{P\'eres}},
  \bibinfo{journal}{Phys. Rev. Lett.}
  \textbf{\bibinfo{volume}{86}}(\bibinfo{number}{20}), \bibinfo{pages}{4676}
  (\bibinfo{year}{2001}).

\bibitem[{\citenamefont{Gu\'eron et~al.}(1999)\citenamefont{Gu\'eron, Deshmukh,
  Myers, and Ralph}}]{gueron1999}
\bibinfo{author}{\bibfnamefont{S.}~\bibnamefont{Gu\'eron}},
  \bibinfo{author}{\bibfnamefont{M.~M.} \bibnamefont{Deshmukh}},
  \bibinfo{author}{\bibfnamefont{E.~B.} \bibnamefont{Myers}}, \bibnamefont{and}
  \bibinfo{author}{\bibfnamefont{D.~C.} \bibnamefont{Ralph}},
  \bibinfo{journal}{Phys. Rev. Lett.}
  \textbf{\bibinfo{volume}{83}}(\bibinfo{number}{20}), \bibinfo{pages}{4148}
  (\bibinfo{year}{1999}).

\bibitem[{\citenamefont{Deshmukh et~al.}(2001)\citenamefont{Deshmukh, Kleff,
  Gu\'eron, Bonnet, Pasupathy, von Delft, and Ralph}}]{deshmukh2001}
\bibinfo{author}{\bibfnamefont{M.~M.} \bibnamefont{Deshmukh}},
  \bibinfo{author}{\bibfnamefont{S.}~\bibnamefont{Kleff}},
  \bibinfo{author}{\bibfnamefont{S.}~\bibnamefont{Gu\'eron}},
  \bibinfo{author}{\bibfnamefont{E.}~\bibnamefont{Bonnet}},
  \bibinfo{author}{\bibfnamefont{A.~N.} \bibnamefont{Pasupathy}},
  \bibinfo{author}{\bibfnamefont{J.}~\bibnamefont{von Delft}},
  \bibnamefont{and} \bibinfo{author}{\bibfnamefont{D.~C.} \bibnamefont{Ralph}},
  \bibinfo{journal}{Phys. Rev. Lett.}
  \textbf{\bibinfo{volume}{87}}(\bibinfo{number}{22}), \bibinfo{pages}{226801}
  (\bibinfo{year}{2001}).

\bibitem[{\citenamefont{Canali and MacDonald}(2000)}]{cmc_ahm2000prl}
\bibinfo{author}{\bibfnamefont{C.~M.} \bibnamefont{Canali}} \bibnamefont{and}
  \bibinfo{author}{\bibfnamefont{A.~H.} \bibnamefont{MacDonald}},
  \bibinfo{journal}{Phys. Rev. Lett.}
  \textbf{\bibinfo{volume}{85}}(\bibinfo{number}{26}), \bibinfo{pages}{5623}
  (\bibinfo{year}{2000}).

\bibitem[{\citenamefont{Papaconstantopoulos}(1986)}]{papaconstantopoulos}
\bibinfo{author}{\bibfnamefont{D.~A.} \bibnamefont{Papaconstantopoulos}},
  \emph{\bibinfo{title}{Handbook of the Band Structure of Elemental Solids}}
  (\bibinfo{publisher}{Plenum, New York}, \bibinfo{year}{1986}).

\bibitem[{\citenamefont{Skomski and Coey}(1999)}]{skomski}
\bibinfo{author}{\bibfnamefont{R.}~\bibnamefont{Skomski}} \bibnamefont{and}
  \bibinfo{author}{\bibfnamefont{J.~M.~D.} \bibnamefont{Coey}},
  \emph{\bibinfo{title}{Permanent Magnetism}} (\bibinfo{publisher}{Institute of
  Physics}, \bibinfo{address}{Bristol}, \bibinfo{year}{1999}).

\bibitem[{\citenamefont{Fradkin}(1991)}]{fradkin}
\bibinfo{author}{\bibfnamefont{E.}~\bibnamefont{Fradkin}},
  \emph{\bibinfo{title}{Field Theories of Condensed Matter Systems}}
  (\bibinfo{publisher}{Addison-Wesley}, \bibinfo{address}{Redwood City},
  \bibinfo{year}{1991}).

\bibitem[{\citenamefont{Canali et~al.}(2002)\citenamefont{Canali, Cehovin, and
  MacDonald}}]{cmc_ac_ahm2002pap4}
\bibinfo{author}{\bibfnamefont{C.~M.} \bibnamefont{Canali}},
  \bibinfo{author}{\bibfnamefont{A.}~\bibnamefont{Cehovin}}, \bibnamefont{and}
  \bibinfo{author}{\bibfnamefont{A.~H.} \bibnamefont{MacDonald}},
  \emph{\bibinfo{title}{Collective excitations in ferromagnetic metal
  nanoparticles}} (\bibinfo{year}{2002}), \bibinfo{note}{unpublished}.

\bibitem[{\citenamefont{Andropov et~al.}(1996)\citenamefont{Andropov,
  Katsnelson, Harmon, van Schilfgaarde, and Kusnezov}}]{andropov1996}
\bibinfo{author}{\bibfnamefont{V.~P.} \bibnamefont{Andropov}},
  \bibinfo{author}{\bibfnamefont{M.~I.} \bibnamefont{Katsnelson}},
  \bibinfo{author}{\bibfnamefont{B.~N.} \bibnamefont{Harmon}},
  \bibinfo{author}{\bibfnamefont{M.}~\bibnamefont{van Schilfgaarde}},
  \bibnamefont{and} \bibinfo{author}{\bibfnamefont{D.}~\bibnamefont{Kusnezov}},
  \bibinfo{journal}{Phys. Rev. B}
  \textbf{\bibinfo{volume}{54}}(\bibinfo{number}{2}), \bibinfo{pages}{1019}
  (\bibinfo{year}{1996}).

\bibitem[{\citenamefont{Uhl et~al.}(1994)\citenamefont{Uhl, Sandratski, and
  Kuber}}]{uhl1994}
\bibinfo{author}{\bibfnamefont{M.}~\bibnamefont{Uhl}},
  \bibinfo{author}{\bibfnamefont{L.~M.} \bibnamefont{Sandratski}},
  \bibnamefont{and} \bibinfo{author}{\bibfnamefont{J.}~\bibnamefont{Kuber}},
  \bibinfo{journal}{Phys. Rev. B}
  \textbf{\bibinfo{volume}{54}}(\bibinfo{number}{20}), \bibinfo{pages}{291}
  (\bibinfo{year}{1994}).

\bibitem[{\citenamefont{Resta}(2000)}]{resta2000}
\bibinfo{author}{\bibfnamefont{R.}~\bibnamefont{Resta}}, \bibinfo{journal}{J.
  Phys.: Condens. Matter} \textbf{\bibinfo{volume}{12}}, \bibinfo{pages}{107}
  (\bibinfo{year}{2000}).

\bibitem[{\citenamefont{Niu and Kleinman}(1998)}]{niu1998}
\bibinfo{author}{\bibfnamefont{Q.}~\bibnamefont{Niu}} \bibnamefont{and}
  \bibinfo{author}{\bibfnamefont{L.}~\bibnamefont{Kleinman}},
  \bibinfo{journal}{Phys. Rev. Lett}
  \textbf{\bibinfo{volume}{80}}(\bibinfo{number}{10}), \bibinfo{pages}{2205}
  (\bibinfo{year}{1998}).

\bibitem[{\citenamefont{Niu et~al.}(1999)\citenamefont{Niu, Wang, Kleinman,
  Liu, Nicholson, and Stocks}}]{niu1999}
\bibinfo{author}{\bibfnamefont{Q.}~\bibnamefont{Niu}},
  \bibinfo{author}{\bibfnamefont{X.}~\bibnamefont{Wang}},
  \bibinfo{author}{\bibfnamefont{L.}~\bibnamefont{Kleinman}},
  \bibinfo{author}{\bibfnamefont{W.~M.} \bibnamefont{Liu}},
  \bibinfo{author}{\bibfnamefont{D.~M.~C.} \bibnamefont{Nicholson}},
  \bibnamefont{and} \bibinfo{author}{\bibfnamefont{G.~M.}
  \bibnamefont{Stocks}}, \bibinfo{journal}{Phys. Rev. Lett}
  \textbf{\bibinfo{volume}{83}}(\bibinfo{number}{1}), \bibinfo{pages}{207}
  (\bibinfo{year}{1999}).

\bibitem[{\citenamefont{Bylander et~al.}(2000)\citenamefont{Bylander, Niu, and
  Kleinman}}]{bylander2000}
\bibinfo{author}{\bibfnamefont{D.~M.} \bibnamefont{Bylander}},
  \bibinfo{author}{\bibfnamefont{Q.}~\bibnamefont{Niu}}, \bibnamefont{and}
  \bibinfo{author}{\bibfnamefont{L.}~\bibnamefont{Kleinman}},
  \bibinfo{journal}{Phys. Rev. B}
  \textbf{\bibinfo{volume}{61}}(\bibinfo{number}{18}), \bibinfo{pages}{11875}
  (\bibinfo{year}{2000}).

\bibitem[{\citenamefont{Gebauer and Baroni}(2000)}]{gebauer2000}
\bibinfo{author}{\bibfnamefont{R.}~\bibnamefont{Gebauer}} \bibnamefont{and}
  \bibinfo{author}{\bibfnamefont{S.}~\bibnamefont{Baroni}},
  \bibinfo{journal}{Phys. Rev. B}
  \textbf{\bibinfo{volume}{61}}(\bibinfo{number}{10}), \bibinfo{pages}{6459}
  (\bibinfo{year}{2000}).

\bibitem[{\citenamefont{MacDonald and Canali}(2001)}]{ahm_cmc2001ssc}
\bibinfo{author}{\bibfnamefont{A.~H.} \bibnamefont{MacDonald}}
  \bibnamefont{and} \bibinfo{author}{\bibfnamefont{C.~M.}
  \bibnamefont{Canali}}, \bibinfo{journal}{Solid State Comm.}
  \textbf{\bibinfo{volume}{119}}, \bibinfo{pages}{253} (\bibinfo{year}{2001}).

\bibitem[{\citenamefont{Cehovin et~al.}(2002)\citenamefont{Cehovin, Canali, and
  MacDonald}}]{ac_cmc_ahm2002}
\bibinfo{author}{\bibfnamefont{A.}~\bibnamefont{Cehovin}},
  \bibinfo{author}{\bibfnamefont{C.~M.} \bibnamefont{Canali}},
  \bibnamefont{and} \bibinfo{author}{\bibfnamefont{A.~H.}
  \bibnamefont{MacDonald}}, \bibinfo{journal}{Phys. Rev. B}
  \textbf{\bibinfo{volume}{66}}(\bibinfo{number}{10}), \bibinfo{pages}{94430}
  (\bibinfo{year}{2002}).

\bibitem[{\citenamefont{Auerbach}(1994)}]{auerbach}
\bibinfo{author}{\bibfnamefont{A.}~\bibnamefont{Auerbach}},
  \emph{\bibinfo{title}{Interacting electrons and quantum magnetism}}
  (\bibinfo{publisher}{Springer-Verlag}, \bibinfo{address}{New York},
  \bibinfo{year}{1994}).

\bibitem[{\citenamefont{Simon}(1983)}]{simon1983}
\bibinfo{author}{\bibfnamefont{B.}~\bibnamefont{Simon}},
  \bibinfo{journal}{Phys. Rev. Lett.}
  \textbf{\bibinfo{volume}{51}}(\bibinfo{number}{24}), \bibinfo{pages}{2167}
  (\bibinfo{year}{1983}).

\bibitem[{\citenamefont{Thouless et~al.}(1982)\citenamefont{Thouless,
  Nightingale, and den Nijs}}]{tkn2}
\bibinfo{author}{\bibfnamefont{D.~J.} \bibnamefont{Thouless}},
  \bibinfo{author}{\bibfnamefont{M.~K. M.~P.} \bibnamefont{Nightingale}},
  \bibnamefont{and} \bibinfo{author}{\bibfnamefont{M.}~\bibnamefont{den Nijs}},
  \bibinfo{journal}{Phys. Rev. Lett.}
  \textbf{\bibinfo{volume}{49}}(\bibinfo{number}{6}), \bibinfo{pages}{405}
  (\bibinfo{year}{1982}).

\bibitem[{\citenamefont{Avron et~al.}(1983)\citenamefont{Avron, Seiler, and
  Simon}}]{avron1983}
\bibinfo{author}{\bibfnamefont{J.~E.} \bibnamefont{Avron}},
  \bibinfo{author}{\bibfnamefont{R.}~\bibnamefont{Seiler}}, \bibnamefont{and}
  \bibinfo{author}{\bibfnamefont{B.}~\bibnamefont{Simon}},
  \bibinfo{journal}{Phys. Rev. Lett.}
  \textbf{\bibinfo{volume}{51}}(\bibinfo{number}{1}), \bibinfo{pages}{51}
  (\bibinfo{year}{1983}).

\bibitem[{\citenamefont{Kohmoto}(1984)}]{kohmoto1985}
\bibinfo{author}{\bibfnamefont{M.}~\bibnamefont{Kohmoto}},
  \bibinfo{journal}{Ann. of Phys.} \textbf{\bibinfo{volume}{160}},
  \bibinfo{pages}{343} (\bibinfo{year}{1984}).

\end{thebibliography}

\end{document}